\documentclass[conference]{IEEEtran}
\IEEEoverridecommandlockouts
\usepackage{hyperref}
\usepackage{flushend}
\usepackage{cite}
\usepackage{amsmath,amssymb,amsfonts}
\usepackage{algorithmic}
\usepackage{graphicx}
\usepackage{textcomp}
\usepackage{xcolor}
\def\BibTeX{{\rm B\kern-.05em{\sc i\kern-.025em b}\kern-.08em
    T\kern-.1667em\lower.7ex\hbox{E}\kern-.125emX}}
\begin{document}

\title{Next-Generation Local Time Stepping for the ADER-DG Finite Element Method
\thanks{The authors acknowledge the Large-Scale Community Partnership ``SCEC Earthquake Modeling, Ground Motion, and Hazard Simulations'' at the Texas Advanced Computing Center (TACC) at The University of Texas at Austin for providing HPC resources that have contributed to the research results reported within this manuscript. This work was supported through the project ``Hocheffiziente und Flexible Deep Learning-Bausteine für Arm- und Power-Prozessoren'', funded by the Carl Zeiss Foundation.}
}

\author{\IEEEauthorblockN{1\textsuperscript{st} Alexander Breuer}
\IEEEauthorblockA{
\textit{Friedrich Schiller University Jena}\\
Jena, Germany \\
alex.breuer@uni-jena.de}
\and
\IEEEauthorblockN{2\textsuperscript{nd} Alexander Heinecke}
\IEEEauthorblockA{
\textit{Intel Corporation}\\
Santa Clara, USA \\
alexander.heinecke@intel.com}
}

\maketitle

\begin{abstract}
High-frequency ground motion simulations pose a grand challenge in computational seismology.
Two main factors drive this challenge.
First, to account for higher frequencies, we have to extend our numerical models, e.g., by considering anelasticity, or by including mountain topography.
Second, even if we were able to keep our models unchanged, simply doubling the frequency content of a seismic wave propagation solver requires a sixteen-fold increase in computational resources due to the used four-dimensional space-time domains.

This work presents the Extreme Scale Discontinuous Galerkin Environment (EDGE) in the context of high-frequency ground motion simulations.
Our presented enhancements cover the entire spectrum of the unstructured finite element solver.
This includes the incorporation of anelasticity, the introduction of a next-generation clustered local time stepping scheme, and the introduction of a completely revised communication scheme.
We close the modeling and simulation loop by presenting our new and rich preprocessing, which drives the high problem-awareness and numerical efficiency of the core solver.

In summary, the presented work allows us to conduct large scale high-frequency ground motion simulations efficiently, routinely and conveniently.
The soundness of our work is underlined by a set of high-frequency verification runs using a realistic setting.
We conclude the presentation by studying EDGE's combined algorithmic and computational efficiency in a demanding setup of the 2014 $\text{M}_\text{w}$ 5.1 La Habra earthquake.
Our results are compelling and show an improved time-to-solution by over 10$\times$ while scaling strongly from 256 to 1,536 nodes of the Frontera supercomputer with a parallel efficiency of over 95\%.
\end{abstract}


\section{2014 Mw 5.1 La Habra Earthquake}
\label{ch:la_habra}
The presented work uses the Extreme Scale Discontinuous Galerkin Environment (EDGE) to tackle the grand challenge of high-frequency ground motion simulations.
A series of simulations of the 2014 $\text{M}_\text{w}$ 5.1 La Habra earthquake guided and accompanied our developments.
\begin{figure}[htbp]
  \centerline{\includegraphics[width=\linewidth]{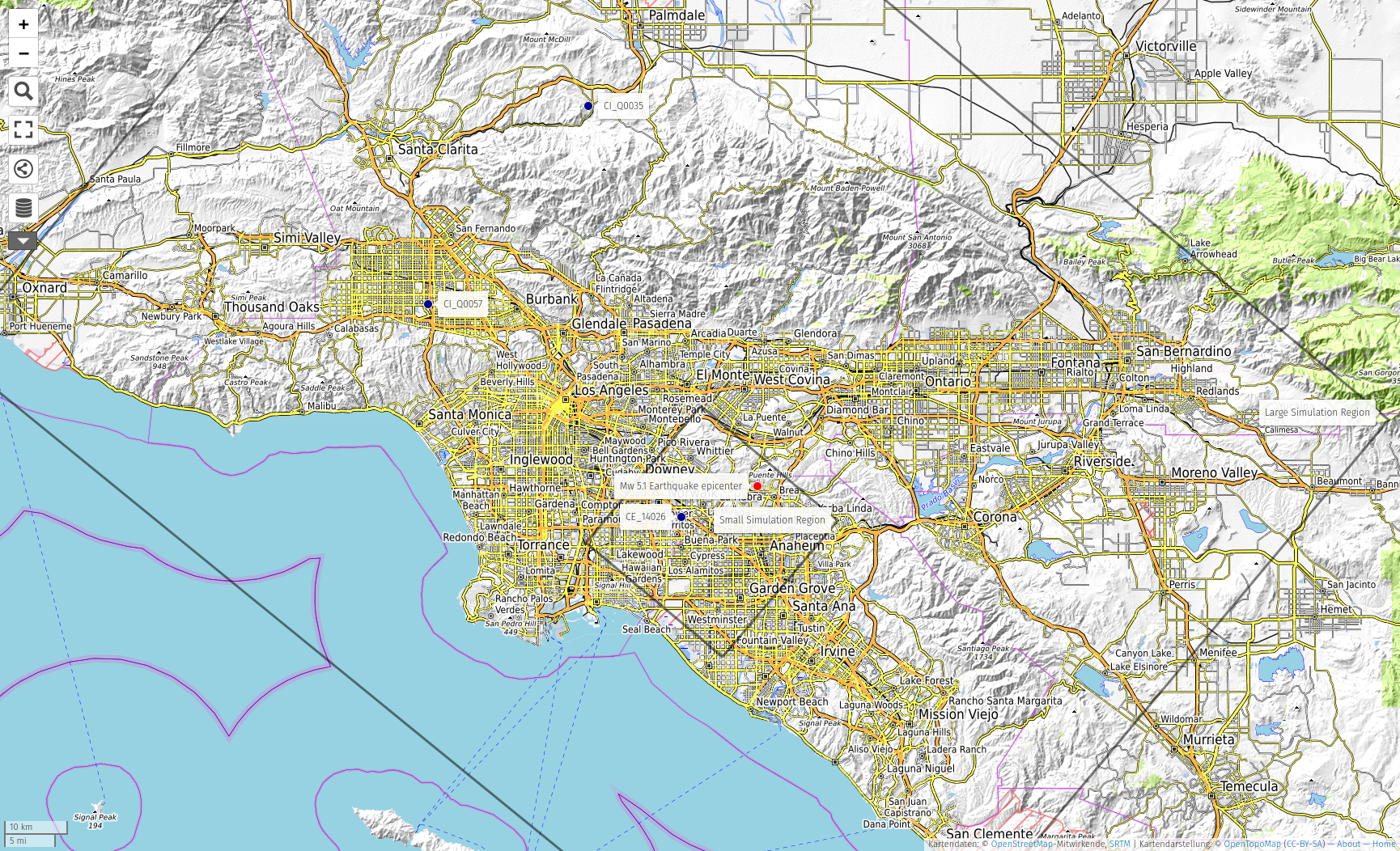}}
  \caption{Study area of the Southern California Earthquake Center's High Frequency project, simulating the 2014 $\text{M}_\text{w}$ 5.1 La Habra, California earthquake. Shown are the ``small'' domain through the inner box and partially the ``large'' domain through the outer box. Additionally, the locations of the earthquake's epicenter and three stations are given. The screenshot was obtained from \url{http://u.osmfr.org/m/560152/}.}
  \label{fig:lahbra_map}
\end{figure}
The High-Frequency (High-F) ground motion verification project of the Southern California Earthquake Center built the initial umbrella of the conducted runs.
High-F specifies inputs for the solvers which are used by the participating modelers:
\begin{itemize}
  \item the modeling assumptions, i.e, anelastic attenuation with a frequency-independent Q-definition;
  \item the targeted frequency content of the simulations, i.e, requiring results which are accurate up to 5\,Hz;
  \item the temporal and spatial extent of the simulations, a map showing High-F's ``small'' and ``large'' domains is given in Fig.\,\ref{fig:lahbra_map};
  \item the used seismic velocity model (CVM-S4.26.M01) with a set of parameter constraints;
  \item the kinematic description of the assumed earthquake rupture; and
  \item the set of seismic stations for which the synthetic seismograms are compared.
\end{itemize}
The goal of the High-F verification effort is a high agreement of the synthetic seismograms when using diverse solvers but the same input.
The project's challenges are driven by the high complexity of the targeted simulations and the high computational demands of the individual forward simulations.
\begin{figure}[htbp]
  \centerline{\includegraphics[width=\linewidth]{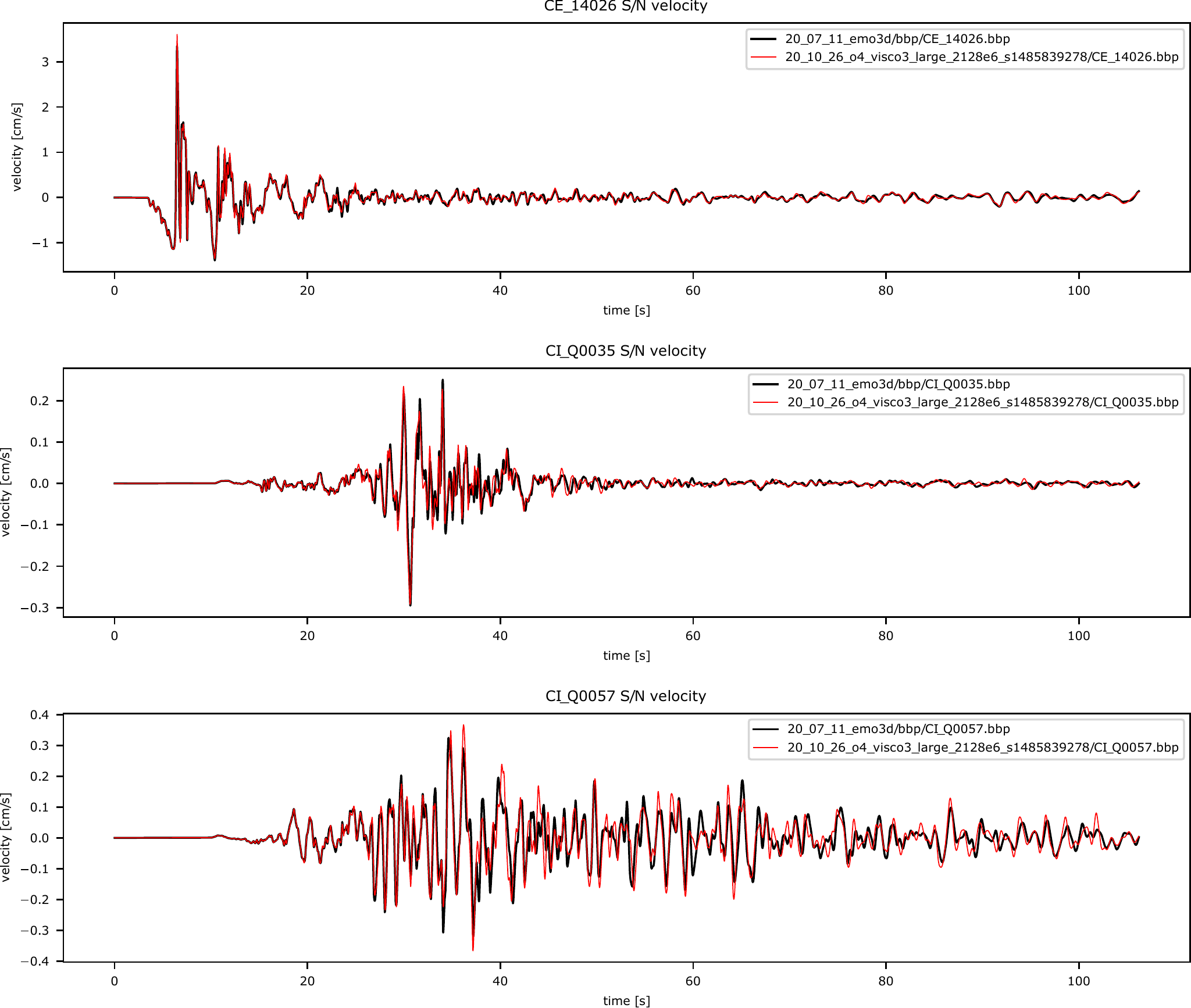}}
  \caption{Comparison of EDGE's South-North velocity component (red) to EMO3D's High-F solution (black). Shown are synthetic seismograms for the three stations depicted in Fig.\,\ref{fig:lahbra_map}. The seismograms were low-pass filtered at 5Hz. EDGE's respective ground motion simulation harnessed 1,536 nodes of the Frontera machine for a total of 48 hours to advance the used 2.1 billion tetrahedral elements in time.}
  \label{fig:verification}
\end{figure}
An exemplary result of this work is given in Fig.\,\ref{fig:verification}.
We observe an excellent agreement of EDGE and the finite-difference solver EMO3D for the South-North velocity component of the three stations depicted in Fig\,\ref{fig:lahbra_map}.

The work presented in this manuscript enabled the solver EDGE for High-F's demanding verification setup.
Additionally, we harnessed the solver's unique capabilities to further increase the complexity of the simulations.
Two key model extensions, posing major obstacles for many earthquake packages, are discussed in detail throughout this manuscript: the introduction of topography and the utilization of problem-aware meshes.
\begin{figure}[htbp]
  \centerline{\includegraphics[width=\linewidth]{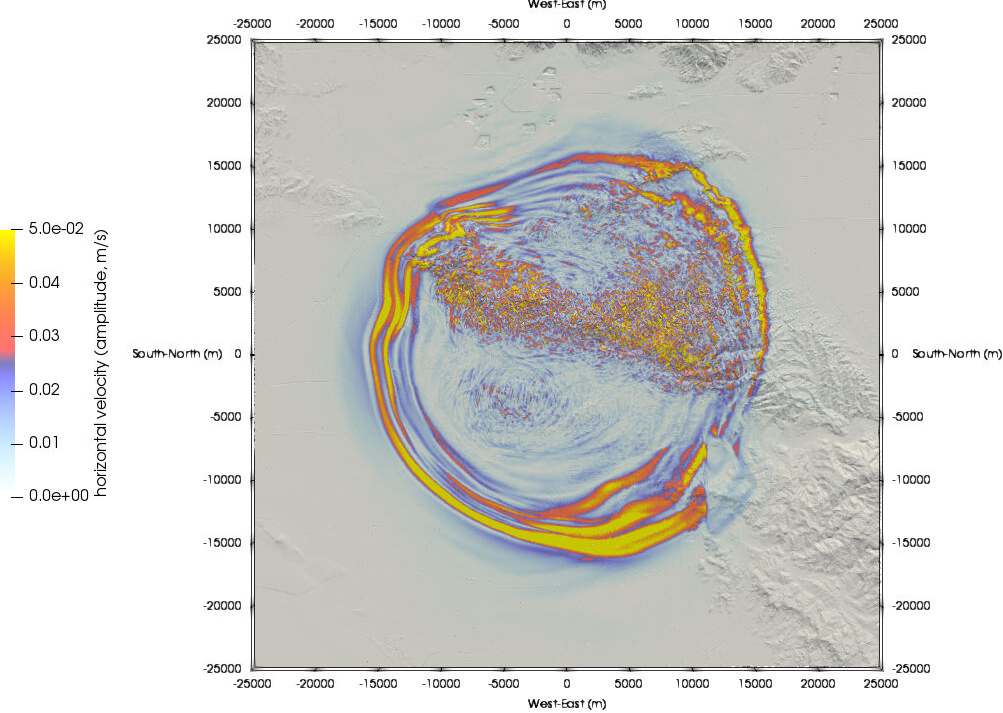}}
  \caption{Visualization of the seismic wave field for a simulation of the 2014 $\text{M}_\text{w}$ 5.1 La Habra earthquake. Shown are the amplitudes of the horizontal particle velocities after seven seconds of simulated time. The West-East and South-North coordinates describe the offset w.r.t. the epicenter.}
  \label{fig:edge_wf}
\end{figure}
An exemplary wave field visualization of a respective simulation for the ``small'' High-F domain is given in Fig.\,\ref{fig:edge_wf}.
In summary, our work makes the following contributions:
\begin{itemize}
  \item the incorporation of the anelastic wave equations into the solver EDGE;
  \item the introduction of a new local time stepping scheme for highly efficient simulations with anelastic attenuation;
  \item the introduction of a new communication scheme minimizing the pressure on the memory and network; and
  \item the introduction of an end-to-end preprocessing pipeline which enables efficient and large scale high-frequency ground motion simulations.
\end{itemize}

\section{Related Work}
Ground motion simulations represent an established pillar of computational seismology.
Their flexibility allows us to gain insight into the earthquake process \cite{ulrich2019dynamic, rodgers2018broadband, olsen2006strong, aagaard2008ground}, quantify seismic hazard \cite{graves2011cybershake}, or invert for important earthquake parameters, e.g., seismic velocities \cite{modrak2018seisflows, lee2014full}.
The earthquake engineering band requires high frequencies beyond 10\,Hz, pushing the limits of simulation software.
This requirement is accompanied by very high computational demands.
These challenges are tackled by variety of multidisciplinary efforts which increase the modeling sophistication \cite{day2001memory, kristek2003seismic, oreilly2021high, roten2014expected, wollherr2018off}, enhance the algorithmic efficiency \cite{nie2017fourth, rietmann2015load, breuer2016petascale}, and enable the solvers for the largest available supercomputers \cite{fu20179, ichimura2015implicit, heinecke14petascale, breuer14sustained, cui2010scalable}.

EDGE \cite{breuer2017edge, breuer2019petaflop}, the solver enhanced through this work, uses the ADER-DG finite element method \cite{dumbser2006three, kaser2007arbitrary, dumbser2007arbitrary}.
ADER-DG is also used in the seismic wave propagation software SeisSol \cite{uphoff2017extreme, heinecke14petascale}.
Both solvers harness the same numerical discretization for the presented settings but rely on different algorithmic formulations.
As a result, the two software packages are completely independent from each other and do not share a single line of code.

EDGE's recent developments especially focus on ensembles of wave propagation simulations in realistic media using kinematic earthquake sources.
SeisSol emphasizes the use of nonlinear dynamic rupture sources and the coupling to Tsunami simulations \cite{id2703, uphoff2017extreme}.
Both EDGE and SeisSol have been extensively optimized for large-scale ground motion simulations.
An efficient and scalable clustered local time stepping scheme for ADER-DG, applied to the elastic wave equations, has been introduced to SeisSol as part of \cite{breuer2016petascale}.
The work described in this manuscript introduces a new next-generation local time stepping scheme to EDGE.
Computationally, both software packages rely heavily on the library LIBXSMM for high-performing small-matrix kernels on a large variety of CPU architectures \cite{heinecke2016libxsmm, georganas21tpp}.
Additionally, SeisSol has been ported to GPUs with an obtained high performance for global time stepping but a mediocre computational efficiency for its crucial local time stepping feature \cite{dorozhinskii2021seissol}.
Current developments of EDGE and LIBXSMM are targeting CPU-native accelerations, e.g., tensor operations on Intel's KnightsMill processor \cite{heinecke2019tensor} or recent CPU-native matrix units \cite{georganas21tpp}.

\section{ADER-DG Finite Element Method}
\label{ch:ader_dg}
The Arbitrary high-order DERivatives (ADER) Discontinuous Galerkin (DG) finite element software EDGE solves the three-dimensional anelastic wave equations.
The equations may be formulated as a linear hyperbolic system with variable coefficients \cite{kaser2007arbitrary}:
\begin{equation}
  q_t + \hat{A} q_x + \hat{B} q_y + \hat{C} q_z = \hat{E} q.
  \label{eq:hyp_sys}
\end{equation}
$t$ is time.
$x$, $y$ and $z$ are the three directions of the Cartesian coordinate system.
The variable vector $q(\vec{x}, t) = [q^e, q^a]^T$ is split into an elastic part $q^e$ and an anelastic part $q^a$. $q^e = ( \sigma_{xx}, \sigma_{yy}, \sigma_{zz}, \sigma_{xy}, \sigma_{yz}, \sigma_{xz}, u, v, w )$ contains the three normal stress components $\sigma_{xx}$, $\sigma_{yy}$ and $\sigma_{zz}$, the shear stresses $\sigma_{xy}$, $\sigma_{yz}$ and $\sigma_{xz}$, and the particle velocities in $x$-, $y$- and $z$-direction given as $u$, $v$ and $v$.
The anelastic part $q^a = ( \vartheta^1_{xx}, \vartheta^1_{yy}, \vartheta^1_{zz}, \vartheta^1_{xy}, \vartheta^1_{yz}, \vartheta^1_{xz}, \ldots, \vartheta^m_{xx}, \vartheta^m_{yy}, \vartheta^m_{zz}, \vartheta^m_{xy}, \vartheta^m_{yz}, \vartheta^m_{xz} )$ comprises a set of $N^a(m) = 6m$ memory variables where $m$ is the number of relaxation mechanisms. Typically, we use three mechanisms for our simulations, meaning that the variable vector $q$ has a total of $N^q = 9 + N^a(3) = 27$ entries.
The three Jacobian matrices $\hat{A}$,  $\hat{B}$ and  $\hat{C}$ are sparse and show symmetry which we exploit heavily when formulating the discrete form:
\begin{equation}
  \hat{A} =
  \begin{bmatrix}
    A^e & 0 \\
    A^a & 0
  \end{bmatrix},
  \hat{B} =
  \begin{bmatrix}
    B^e & 0 \\
    B^a & 0
  \end{bmatrix},
  \hat{C} =
  \begin{bmatrix}
    C^e & 0 \\
    C^a & 0
  \end{bmatrix} \in \mathbb{R}^{N^q \times N^q}.
\end{equation}
$A^e, B^e, C^e \in \mathbb{R}^{9\times9}$ are the Jacobians of the elastic part as given in \cite{dumbser2006three}.
The matrices $A^a, B^a, C^a \in \mathbb{R}^{6 \times 9}$ carry the anelastic part as described in \cite{kaser2007arbitrary}.
The matrix $\hat{E}$, with the block-structured part $E = [E_1, \ldots, E_m] \in \mathbb{R}^{9 \times 6m}$ and the diagonal part $E'$, couples the anelastic and elastic parts of the system:
\begin{equation}
  \hat{E} =
  \begin{bmatrix}
    0 & E \\
    0 & E'
  \end{bmatrix} \in \mathbb{R}^{N^q \times N^q}.
\end{equation}

\subsection{Discrete Formulation}
Application of the ADER-DG machinery to Eq.\,\eqref{eq:hyp_sys} results in the fully discrete formulation.
We use conforming unstructured tetrahedral meshes for our spatial discretization.
$K \in \mathbb{N}^+$ is the total number of tetrahedrons in a mesh.
For each tetrahedron $k$ with $1 \le k \le K$, we define the modal coefficients of a distinct set of $\mathcal{B}(\mathcal{O})$ polynomial basis functions defined in terms of a unique reference element.
Here, our basis is derived through the tetrahedral expansion outlined in \cite{karniadakis2013spectral}.
Typically, we use order $\mathcal{O}=4$ or $\mathcal{O}=5$ accurate space-approximations, resulting in $\mathcal{B}(4)=20$ or $\mathcal{B}(5)=35$ basis functions.
In summary, the  Degrees Of Freedom (DOFs) $Q_k = [Q_k^e, Q_k^a]^T \in \mathbb{R}^{N^q \times \mathcal{B}}$ for a tetrahedral element $k$ discretize the variables.
$Q_k^e \in \mathbb{R}^{9 \times \mathcal{B}}$ are the DOFs corresponding to the elastic variables.
Analogously, $Q_k^a = [Q_k^{a,1}, \ldots, Q_k^{a,m}]^T \in \mathbb{R}^{ 6m \times \mathcal{B}}$ corresponds to the anelastic variables where $Q_k^{a,l} \in \mathbb{R}^{ 6 \times \mathcal{B}}$ reflects those of a single relaxation mechanism with $1 \le l \le m$.
\paragraph{Time Kernel}
Our time predictor is given by the ADER-scheme.
The DOFs of element $k$ are integrated over the interval $[t_0, t_0 + \Delta t]$ by a Taylor approximation in terms of the time derivatives $\partial^j/\partial t^j Q_k(t_0)$ about the expansion point $t_0$:
\begin{equation}
  \label{eq:ader_taylor}
  \begin{aligned}
      &\mathcal{T}_k(t_0, {\Delta t}) = [\mathcal{T}_k^e, \mathcal{T}_k^a]^T = \\
      &\int_{t_0}^{t_0+\Delta t} Q_{k}(t_0,t) \,\mathrm{d}t
    = \sum_{d=0}^{\mathcal{O}-1} \frac{\left(\Delta t \right)^{d+1} }{(d+1)!} \cdot \frac{\partial^d}{\partial t^d} Q_k(t_0).
  \end{aligned}
\end{equation}
$\mathcal{T}_k^e \in \mathbb{R}^{9 \times \mathcal{B}}$ is the elastic part of the time-integrated DOFs $\mathcal{T}_k \in \mathbb{R}^{N^q \times \mathcal{B}}$, and $\mathcal{T}_k^a \in \mathbb{R}^{6m \times \mathcal{B}}$ the anelastic part.
We obtain the time derivatives through the Cauchy Kowalevski procedure which repeatedly uses Eq.\,\eqref{eq:hyp_sys} to replace time derivatives with space derivatives:
\begin{equation}
  \label{eq:ader_derivative}
  \begin{aligned}
    \frac{\partial^{d+1}}{\partial t^{d+1}} Q_k(t_0) =
    \left[
    \frac{\partial^{d+1}}{\partial t^{d+1}} Q^e_k(t_0),
    \frac{\partial^{d+1}}{\partial t^{d+1}} Q^a_k(t_0)
    \right]^T,
  \end{aligned}
\end{equation}

\begin{equation}
  \label{eq:ader_derivative_elastic}
    \frac{\partial^{d+1}}{\partial t^{d+1}} Q^e_k =
    - \sum_{c=1}^3
    \begin{bmatrix}
      \bar{A}^e_{k,c}
      \left(  \frac{\partial^{d}}{\partial t^{d}} Q^e_k  \right)
      \left( K_c \right)^T
    \end{bmatrix}
    + \sum_{l=1}^{m} E^l_k \frac{\partial^{d}}{\partial t^{d}} Q^{a,l}_{k},
\end{equation}
\begin{equation}
  \label{eq:ader_derivative_anelastic}
    \frac{\partial^{d+1}}{\partial t^{d+1}} Q^{a,l}_{k} =
    - \omega_l \sum_{c=1}^3
    \begin{bmatrix}
      \bar{A}^{a}_{k,c}
      \left(  \frac{\partial^{d}}{\partial t^{d}} Q^e_k  \right)
      \left( K_c \right)^T
    \end{bmatrix}
    + \omega_l \frac{\partial^{d}}{\partial t^{d}} Q^{a,l}_{k}.
\end{equation}
$K_c \in \mathbb{R}^{\mathcal{B} \times \mathcal{B}}$ with $c \in \{1,2,3\}$ are the three stiffness matrices, defined in terms of the unique reference element and pre-multiplied by the inverse, diagonal mass matrix in preprocessing.
$\bar{A}^e_{k,c} \in \mathbb{R}^{9\times9}$ are element-local linear combinations of the elastic Jacobians. $\bar{A}^{a}_{k,c} \in \mathbb{R}^{6 \times 9} $ are linear combinations of the Jacobians' anelastic part, where we factorized the relaxation frequencies $\omega_l$ in Eq.\,\eqref{eq:ader_derivative_anelastic}.
Thus, in the actual implementation, we only compute the sum in Eq.\,\eqref{eq:ader_derivative_anelastic} once and simply scale it to obtain the time derivatives corresponding to a single relaxation mechanism $l$.
Similarly, we reuse the intermediate result $(\partial^d / \partial t^d Q_k^e) (K_c)^T$ of the elastic derivative computation in Eq.\,\eqref{eq:ader_derivative_elastic} for the anelastic computations in Eq.\,\eqref{eq:ader_derivative_anelastic}.
We use element $k$'s DOFs $Q_k^{n_k}$  after its $n_k$-th time step at simulation time $t_k^{n_k}$ as expansion point.
Thus, these DOFs build the initial values of the recursive procedure in equations \eqref{eq:ader_derivative_elastic} and \eqref{eq:ader_derivative_anelastic}, i.e., $\partial^0 / \partial t^0 Q_k(t_0) = Q_k^{n_k}$.
\paragraph{Volume Kernel}
The element-local volume kernel operates on the time-integrated DOFs $\mathcal{T}_k$.
As for the time kernel, we split the volume kernel into an elastic part $\mathcal{V}_k^e \in \mathbb{R}^{9 \times \mathcal{B}}$ and an anelastic part $\mathcal{V}_k^a = [\mathcal{V}_k^{a,1}, \ldots, \mathcal{V}_k^{a,m}]^T \in \mathbb{R}^{6m \times \mathcal{B}}$:
\begin{equation}
  \mathcal{V}_k^e = \sum_{c=1}^{3}
  \left[ \bar{A}^{e}_{k,c} \left(  \mathcal{T}_k^e  \right) K_c \right]
  + \sum_{l=1}^{m} E^l_k \mathcal{T}_k^{a,l},
\end{equation}
\begin{equation}
  \mathcal{V}_k^{a,l} = \omega_l \sum_{c=1}^{3}
  \left[ \bar{A}^{a}_{k,c} \left(  \mathcal{T}_k^e  \right) K_c \right]
  - \omega_l \mathcal{T}_k^{a,l}.
\end{equation}
Similarly to the time kernel, we reuse the intermediate result $\left(  \mathcal{T}_k^e  \right) K_c$ of the elastic part $\mathcal{V}_k^e$ when computing the anelastic part $\mathcal{V}_k^{a,l}$.

\paragraph{Surface Kernel}
As the time and volume kernels, the surface kernel is split into an elastic part $\mathcal{S}_k^e = \mathcal{S}_{k}^{e_L} + \mathcal{S}_{k}^{e_N}$  and an anelastic part $\mathcal{S}_k^a = \mathcal{S}_k^{a_L} + \mathcal{S}_k^{a_N} = [ \mathcal{S}_{k}^{a_L,1}+\mathcal{S}_{k}^{a_N,1}, \ldots , \mathcal{S}_{k}^{a_L,m}+\mathcal{S}_{k}^{a_N,m}]^T$.
$\mathcal{S}_{k}^{e_L}$ only uses the elastic time-integrated DOFs $\mathcal{T}^e_k$ of element $k$ itself, whereas $\mathcal{S}_{k}^{e_N}$ uses the elastic ones of $k$'s four face-neighboring tetrahedrons $\mathcal{T}^e_{k_1}$, $\mathcal{T}^e_{k_2}$, $\mathcal{T}^e_{k_3}$ and $\mathcal{T}^e_{k_4}$:
\begin{equation}
  \label{eq:local_elastic}
  \mathcal{S}_{k}^{e_L}
  = \sum_{i=1}^4 \left( \tilde{A}_{k,i}^{e,-} \left( \left( \mathcal{T}^e_k \right) \tilde{F}_i \right) \right) \hat{F}_i,
\end{equation}
\begin{equation}
  \label{eq:neigh_elastic}
  \mathcal{S}_{k}^{e_N}
  = \sum_{i=1}^4 ( \tilde{A}_{k,i}^{e,+}  ( \underbrace{ ( \mathcal{T}^e_{k_i} ) \bar{F}_{j_k(i), h_k(i)}  }_{\text{MPI}} ) ) \hat{F}_i.
\end{equation}
The anelastic part $\mathcal{S}_k^a$ only requires the time-integrated elastic DOFs of element $k$ for $\mathcal{S}_{k}^{a_L}$, and those of the face-neighboring tetrahedrons $k_1$, $k_2$, $k_3$ and $k_4$ for $\mathcal{S}_{k}^{a_N}$:
\begin{equation}
  \label{eq:local_anelastic}
  \mathcal{S}_{k}^{a_L,l}
  = \omega_l \sum_{i=1}^4 \left( \tilde{A}_{k,i}^{a,-} \left( \left( \mathcal{T}^e_k \right) \tilde{F}_i \right) \right) \hat{F}_i,
\end{equation}
\begin{equation}
  \label{eq:neigh_anelastic}
  \mathcal{S}_{k}^{a_N,l}
  = \omega_l \sum_{i=1}^4 ( \tilde{A}_{k,i}^{a,+}  ( \underbrace{ ( \mathcal{T}^e_{k_i} ) \bar{F}_{j_k(i), h_k(i)}  }_{\text{MPI}} ) ) \hat{F}_i.
\end{equation}
The matrices $\tilde{A}_{k,i}^{e,-}, \tilde{A}_{k,i}^{e,+} \in \mathbb{R}^{9 \times 9}$ and $\tilde{A}_{k,i}^{a,-}, \tilde{A}_{k,i}^{a,+} \in \mathbb{R}^{6 \times 9}$ are the element-local flux solvers.
$\tilde{F}_i \in \mathbb{R}^{\mathcal{B} \times \mathcal{F}}$ are the four unique flux-matrices \cite{heinecke2019tensor}. $\hat{F}_i \in \mathbb{R}^{\mathcal{F} \times \mathcal{B}}$ are the four transposed flux matrices, pre-multiplied with the inverse mass matrix in pre-processing \cite{heinecke2019tensor}.
Here, $\mathcal{F}(\mathcal{O})$ gives the number of triangular basis functions, i.e., $\mathcal{F}(5)=15$ for the considered setups in this work.
$\bar{F}_{j_k(i), h_k(i)} \in \mathbb{R}^{\mathcal{B} \times \mathcal{F}}$ with $j_k(i) \in \{1,2,3,4\}$ and $h_k(i) \in \{1, 2, 3\}$ are the 12 unique neighboring flux matrices \cite{heinecke2019tensor}.
Note that our implementation reuses the intermediate results $(\mathcal{T}^e_k)\tilde{F}_i$ and $(\mathcal{T}^e_{k_i})\tilde{F}_{j_k(i),h_k(i)}$ of the elastic part in Eq.\,\eqref{eq:local_elastic} and Eq.\,\eqref{eq:neigh_elastic} when computing the anelastic part of the surface kernel through Eq.\,\eqref{eq:local_anelastic} and Eq\,\eqref{eq:neigh_anelastic}.

\paragraph{Update Scheme}
The final update scheme uses the time-integrated DOFs of element $k$ and its face-neighboring elements in the volume and surface kernels to  perform the element-local timestep $t^{n_k} \rightarrow t^{n_k+1}$:
\begin{equation}
  \label{eq:update}
  Q_k^{n_k+1} = Q_k^{n_k} + \underbrace{ \left[ \mathcal{V}^e, \mathcal{V}^a \right]^T + \left[ S^{e_L}, S^{a_L} \right]^T }_\text{local} + \underbrace{ \left[ S^{e_N}, S^{a_N} \right]^T }_\text{neighboring}.
\end{equation}
As outlined in Eq.\,\eqref{eq:update}, the solver EDGE splits an element $k$'s update into a local step, only requiring data of $k$ itself and a neighboring step, which only requires data of the face-neighboring elements.
We use this splitting to hide communication required for the neighboring updates behind other computation.

\section{Efficient ADER-DG Kernels}
EDGE's implementation of Sec.\,\ref{ch:ader_dg}'s numerics targets best time-to-solution by maximizing the throughput of the performed floating point operations and by maximizing the value of each operation.
Targeting both the throughput and the value of the floating point operations typically requires tradeoffs on both ends for best time-to-solution. 

\subsection{Sparsity and Fused Simulations}
\label{ch:sparse_fused}
A large subset of the used matrices, i.e., $\bar{A}^e_{k,c}$, $\bar{A}^a_{k,c}$, $K_c$, $E_k^l$, $\tilde{F}_i$, $\bar{F}_{j_k(i), h_k(i)}$ and $\hat{F}_i$ are sparse.
Additionally, some matrices, e.g., $K_c$ or $\bar{A}^a_{k,c}$, exhibit block-sparsity.
An illustration of the sparsity patterns for the elastic wave equations and $\mathcal{O}=4$ is given in \cite{heinecke2019tensor}.
The solver EDGE utilizes the matrices' block-sparsity in all studied configurations of this work.
The sparsity patterns are static and the number of involved matrices is small.
Thus, in contrast to \cite{uphoff2016generating}, we simply derived our exploitation of block-sparsity manually.

Additionally, we might exploit the sparsity outside of the matrices' zero-blocks.
This is challenging due to vectorization requirements of modern processors.
EDGE supports running fused simulations \cite{breuer2017edge}, which allow us to inject a multitude of seismic sources into one execution of the solver.
If the number of fused simulations matches the vector width of the considered Instruction Set Architecture (ISA), we are able to vectorize the sparse matrix operations perfectly over the ensemble simulations.
In this work, we study EDGE by 1) running single forward simulations (only block-sparsity is exploited), and 2) by running sixteen concurrent forward simulations (all sparsity is exploited).
Here, using sixteen simulations matches AVX512's vector width for the targeted single precision arithmetic runs on the Frontera machine.

\subsection{Runtime Code-Generation}
As discussed in Sec.\,\ref{ch:ader_dg} and Sec.\,\ref{ch:sparse_fused}, EDGE's ADER-DG kernels rely on dense and sparse linear algebra operations of very limited dimensionality.
The number of rows or columns of the involved matrices is upper-bound by the number of basis functions.
For the considered fifth order scheme, i.e., $\mathcal{O}=5$, we obtain $M,N,K \le \mathcal{B}(5)=35$ when using BLAS-3 parameters for the respective sizes in the matrix-matrix products.

Vendor-tuned and -maintained BLAS libraries are a suboptimal choice for this case.
Typically, these libraries a) do not optimize for the respective odd and small matrix sizes which barely match the underlying hardware SIMD/vector length in case of dense operations, and b) only optimize for extreme sparsity (99+\%) as observed in many established numerical solvers.
However, relying on a compiler to vectorize perfectly, even with pragma-guided vectorization, is a cumbersome task.
This is especially challenging when targeting portability between different platforms, e.g., x86 and AArch64 ISAs, and inter-operability with different compilers, e.g., the Intel compiler, the GNU compiler tool chain, or the LLVM-based clang.

All of these aspects are addressed by LIBXSMM and its Tensor Processing Primitives (TPPs) approach\cite{georganas21tpp}. TPPs can be regarded as a virtual Tensor ISA which is compiled at run-time to allow for best possible mapping of the linear algebra operators to the underlying hardware. TPPs define a compact, yet versatile set of 2D-tensor operators (hence virtual Tensor ISA), which subsequently can be utilized as building-blocks to construct complex operators on high-dimensional tensors.
The TPPs' specification is agnostic to the targeted platform, i.e., the framework it is used in, and the compiler back-end.
Hence a TPP-code is portable, offering significantly better out-of-the-box performance, while allowing potential for a highly optimized, platform-specific implementation.
Today, TPPs outperform state-of-the-art vendor-BLAS-libraries on multiple platforms, including multiple Intel (SSE/AVX/AVX512), AMD and ARM (NEON/SVE) CPUs.

EDGE's support for fused forward simulations can be sped-up by leveraging the sparse nature of the involved matrices, as outlined in Sec.\,\ref{ch:sparse_fused}.
Some TPPs which are targeting sparse deep learning are a good match.
However, we developed and added some changes for the smaller sizes occurring in EDGE, and also contributed an ARM back-end of those.
All of these changes are up-streamed at the time of this writing.
In detail, these improvements target the following two TPPs: a) multiply a sparse matrix with a 3D tensor which represents a  matrix of vectors, and b) multiply a similar 3D tensor with a sparse matrix.
These 3D tensors are naturally created when executing in EDGE's fused forward mode as each scalar DOF of a single simulation corresponds to a vector in the fused setting.
A detailed description of a standalone (non-TPP-version) of this operator targeting the Intel Knights architecture is described in \cite{heinecke2019tensor}.
We extended these TPPs to support arbitrary partial vector processing and added support for ARM processors.

Nevertheless, we can conclude that all TPPs, used in and enhanced for this work, run at 70-90\% of the theoretical peak performance for the studied Cascade Lake processors of Frontera (see Sec.\,\ref{ch:simulations}) out of private caches.
In summary, TPPs applied to HPC allow us to use any C++ compiler and any platform for good performance since compiler readiness, optimization and vectorization pragma and/or flag tuning is no longer required.

\section{Next-Generation Local Time Stepping}
\label{ch:next_gen_lts}
The multi-rate Local Time Stepping (LTS)-scheme introduced in \cite{breuer2016petascale} was designed with the elastic wave equations in mind and follows a buffer-derivative paradigm.
Assume that element $k$ has a time step which is twice as large as that of a face-neighboring element $k_\text{neigh}$: $\Delta t_k = 2 \Delta t_{k_\text{neigh}}$.
Thus, whenever $k$ does one time step, $k_\text{neigh}$ does two.
In this case, $k_\text{neigh}$ would sum two consecutive time-integrated elastic DOFs in a buffer and provide this information to $k$ for its time step.
Element $k$, however, would simply share its time derivatives with $k_\text{neigh}$.
Here, we can evaluate Eq.\,\eqref{eq:ader_taylor} twice for $k_\text{neigh}$'s two time steps.

In the case of the elastic wave equations, we can exploit zero-blocks in the higher time derivatives \cite{breuer14sustained} when communicating time derivatives to neighboring elements through the memory or the network.
However, in the case of the considered anelastic wave equations, the elastic time derivatives in Eq.\,\eqref{eq:ader_derivative_elastic} are coupled through the reactive source to the anelastic time derivatives.
Therefore, we cannot exploit zero-blocks and would have to communicate an excessive amount of data if following \cite{breuer2016petascale}'s approach for the anelastic wave equations.
Specifically, for the considered fifth order of convergence, the five elastic derivatives would require $5 \cdot 9 \cdot 35 = 1,575$ values.
In this work, we introduce a next-generation local time stepping scheme which follows a new paradigm and operates efficiently in the elastic and anelastic case.
Additionally, we present a simple but efficient preprocessing step which improves the algorithmic efficiency of the final scheme.

EDGE's next-generation local time stepping scheme was designed from scratch to accelerate the anelastic wave equations and to reduce the complexity of the implementation.

\subsection{Clustering}
\label{ch:clustering}
Our new clustering approach is a revised version of the rate-2 scheme, described in \cite{breuer2016petascale} and, e.g., used in \cite{id2703, uphoff2017extreme}.
Assume that each element $k$ has a local time step $\Delta t_k^\text{CFL}$, then the minimum time step of all $K$ elements is given as:
\begin{equation}
  \Delta t_{\min}^\text{CFL} = \min_{k=1, \ldots, K} \Delta t_k^\text{CFL}.
\end{equation}
A Global Time Stepping (GTS) scheme advances all elements at this worst-case time step.
GTS will thus lose algorithmic efficiency whenever $\Delta t_k^\text{CFL} > \Delta t_{\min}^\text{CFL}$ for an element $k$.
However, the regularity of GTS w.r.t. the time dimension helps with computational efficiency.
In contrast, advancing each element $k$ at its optimal local time step $\Delta t_k^\text{CFL}$, as done in \cite{dumbser2007arbitrary}, is highly complex and typically results in a low throughput of element updates.
Our clustering aims at a high algorithmic efficiency while maintaining the GTS-throughput.
For this, we define a set of $N_C \ge 1$ time clusters, which cover the following time step intervals with $\lambda \in (0.5, 1]$:
\begin{equation}
  \begin{aligned}
    C_1 = & [ \lambda \cdot \Delta t_{\min}^\text{CFL}, \quad & 2 \lambda \cdot \Delta t_{\min}^\text{CFL} ), \\
    C_2 = & [ 2\lambda \cdot \Delta t_{\min}^\text{CFL}, \quad & 2^2 \lambda \cdot \Delta t_{\min}^\text{CFL} ), \\
    \ldots \\
    C_{N_c} = &[ 2^{N_c-1}\lambda \cdot \Delta t_{\min}^\text{CFL}, & \infty).
  \end{aligned}
\end{equation}
Now, we simply assign each element based on its local time step to the respective cluster.
For example, if element $k$ has the time step $3 \lambda \cdot \Delta t_{\min}^\text{CFL}$, it belongs to cluster $C_2$.
All elements in a specific cluster $C_l$ advance with the respective lower-bound time step, i.e., $2^{l-1} \lambda \cdot \Delta t_{\min}^\text{CFL}$.
Additionally, we follow \cite{breuer2016petascale} to normalize the clustering so that elements in a cluster $C_l$ are only allowed to have neighboring elements in clusters $C_l$, $C_{l-1}$ and $C_{l+1}$.
For example, assume that an element $k$ is initially assigned to $C_3$ but has a neighboring element in $C_1$.
In this case, we would move $k$ from $C_3$ to $C_2$ to satisfy our normalization requirement.
Note, that the normalization circumvents corner cases in the implementation but comes with a negligible loss in algorithmic efficiency, when applied in practice.
In the studied settings of Sec.\,\ref{ch:simulations}, this loss is below 1.5\%.

We observe two major differences of our approach when comparing it to the scheme presented in \cite{breuer2016petascale}.
First, the number of clusters $N_c$ can be set by our users, whereas the scheme in \cite{breuer2016petascale} does not use the open-ended cluster $C_{N_c}$. Instead it generates a sufficient amount of clusters to cover all elements' time steps.
In practice, often times only very few elements have large time steps, meaning that three to five clusters are sufficient for an efficient LTS scheme.
Additionally, as discussed further in Sec.\,\ref{ch:distributed_memory}, a limit on the total number of time clusters mitigates heterogenous memory footprints of partitions in distributed memory settings.

Second, our scheme is limited to rate-2 LTS but allows for the new parameter $\lambda$.
Limiting the scheme to multiples of two when considering the clusters' time intervals is motivated by storage requirements of the anelastic implementation as outlined in Sec.\,\ref{ch:buffers}.
The parameter $\lambda$ is used to optimize the time intervals of the clusters.
For example, assume that the large majority of the elements' time steps is in $(3.0 \Delta t_{\min}^\text{CFL}, 4.0 \Delta t_{\min}^\text{CFL})$.
Using $\lambda=0.75$, we would have $C_1=[0.75 \Delta t_{\min}^\text{CFL}, 1.5 \Delta t_{\min}^\text{CFL})$, $C_2=[1.5 \Delta t_{\min}^\text{CFL}, 3.0 \Delta t_{\min}^\text{CFL})$ and so on.
This means that the large majority of elements advances with $3.0 \Delta t_{\min}^\text{CFL}$ for $\lambda = 0.75$ instead of $2.0 \Delta t_{\min}^\text{CFL}$ as they would with $\lambda=1$.
\begin{figure}[htbp]
  \centerline{\includegraphics[width=1.0\linewidth]{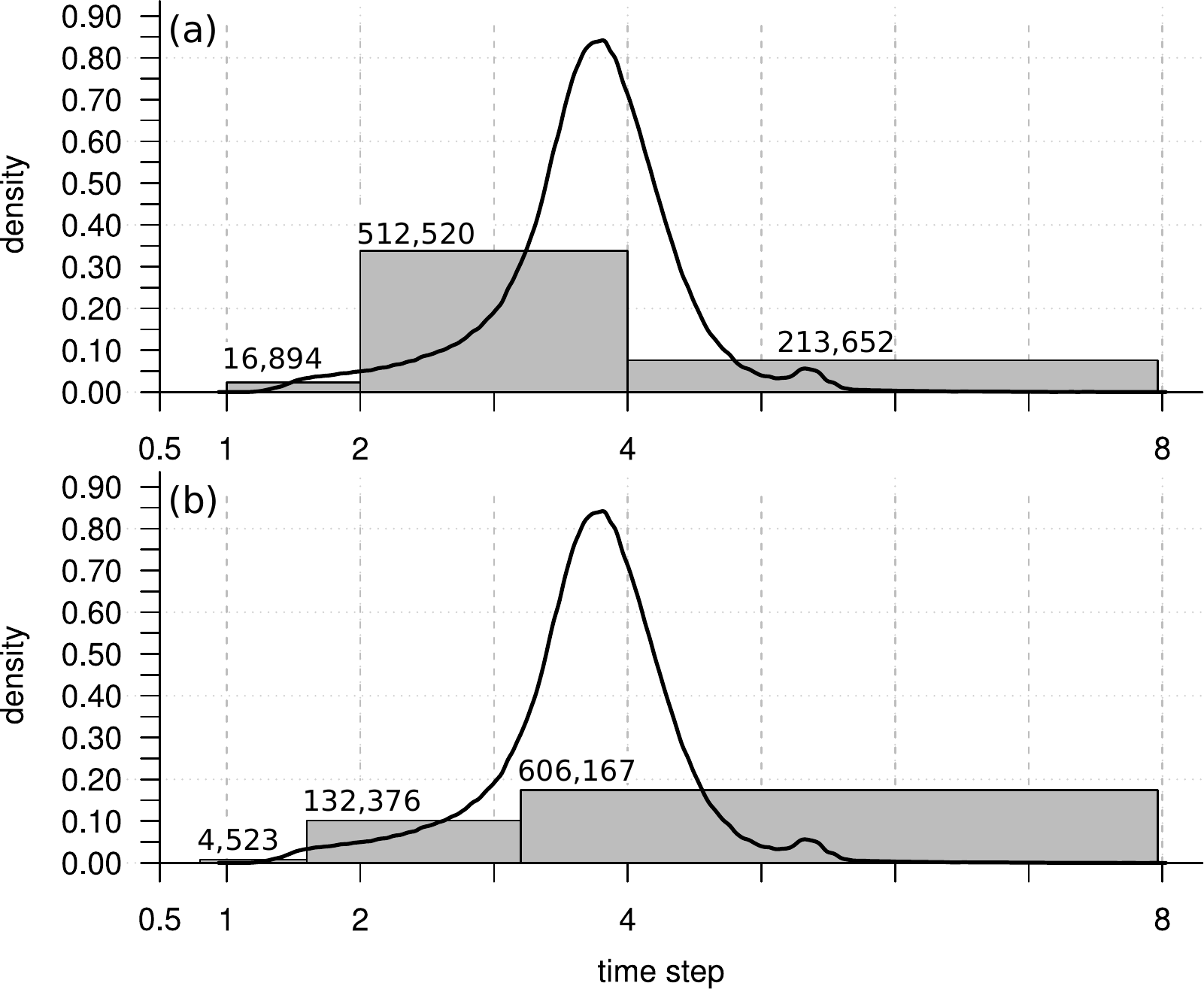}}
  \caption{Illustration of the time step distribution for the studied 743,066-element setup of the LOH.3 benchmark. The solid line shows the time step density of the mesh elements. The gray boxes show the density of the clustering scheme for $\lambda=1.00$ in (a) and $\lambda=0.80$ in (b). On top of each cluster the number of contained elements is given. The time step relative to $\Delta t_{\min}^\text{CFL}$ is given on the x-axis and the element-density on the y-axis.}
  \label{fig:ts_density_loh3}
\end{figure}
In practice, we use a simple and fast preprocessing step which tests possible values of $\lambda$ with an increment of $0.01$.

\begin{figure}[htbp]
  \centerline{\includegraphics[width=1.0\linewidth]{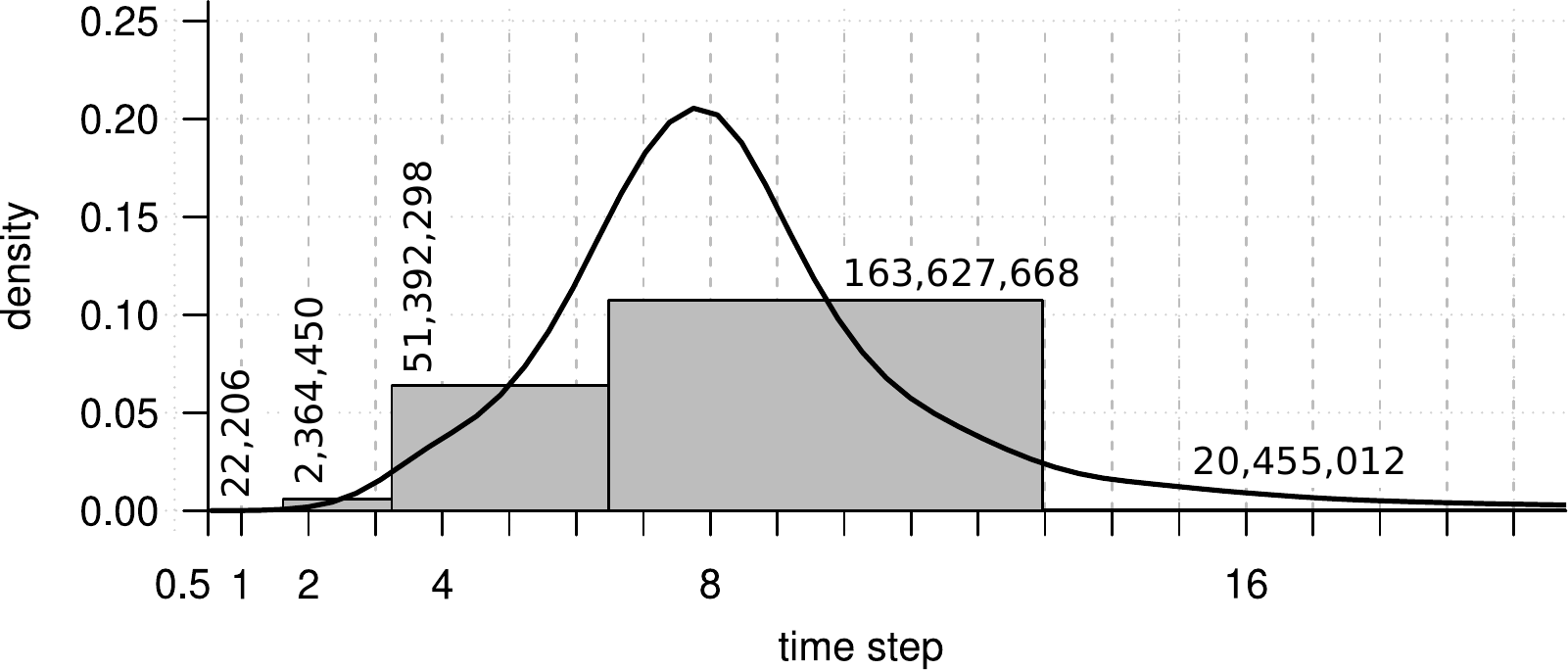}}
  \caption{Illustration of the time step distribution for the 237,861,634-element 2014 $\text{M}_\text{w}$ 5.1 La Habra setting. The solid line shows the time step density of the mesh elements. The gray boxes show the density of the used clustering scheme with $\lambda=0.81$. On top of each cluster the number of contained elements is given. The time step relative to $\Delta t_{\min}^\text{CFL}$ is given on the x-axis and the element-density on the y-axis.}
  \label{fig:ts_density_la_habra}
\end{figure}

Fig.\,\ref{fig:ts_density_loh3} and Fig.\,\ref{fig:ts_density_la_habra} illustrate our clustering for settings used in Sec.\ref{ch:simulations}.
All plots show the elements' time step density as a solid line and the clustering as gray boxes.
The area under the curve sums to 1 respectively.
Additionally, the number of elements in each of the clusters is given on top of the boxes.
Fig.\,\ref{fig:ts_density_loh3}\,(a) shows the clustering of the 743,066-element LOH.3 setup (see Sec.\,\ref{ch:simulations}) for $N_c=3$ and $\lambda=1.0$.
We obtain a theoretical speedup of $2.28\times$ over GTS and observe that cluster $C_2$ has most elements (512,520) and carries most of the computational load (78.5\%).
However, when tuning the parameter $\lambda$, and using $\lambda = 0.80$, we obtain the situation illustrated in Fig.\,\ref{fig:ts_density_loh3}\,(b).
Now, cluster $C_3$ has the majority of elements (606,167) and carries most of the load (68.2\%).
This results in a theoretical $2.67\times$ speedup over GTS, which is a 17.5\% improvement over using $\lambda=1.0$.

Fig.\,\ref{fig:ts_density_la_habra} shows our clustering for a setup of the 2014 $\text{M}_\text{w}$ 5.1 La Habra earthquake.
In this case, we used $N_c=5$ and $\lambda=0.81$, and obtained a theoretical $5.38\times$ speedup over GTS.
Compared to the LOH.3-setting in Fig.\,\ref{fig:ts_density_loh3}, the higher theoretical speedup originates from the larger relative time step (compared to $\Delta t_{\min}^\text{CFL}$) of the bulk of elements.

\subsection{Buffers and Algorithm}
\label{ch:buffers}
In addition to the degrees of freedom $Q_k$ for an element $k$, we introduce the three additional data structures $B^1_k$, $B^2_k$ and $B^3_k$ of size $9 \times \mathcal{B}$.
These data structures are used to store information required by face-neighboring elements with respect to the local LTS-configuration in memory.
For element $k$ at time level $t_k^{n_k}$, after computing the elastic time derivatives (see Eq.\,\eqref{eq:ader_derivative_elastic}), the data structures are used to store the information required by face-neighboring elements in Eq.\,\eqref{eq:neigh_elastic} and Eq.\,\eqref{eq:neigh_anelastic}:
\begin{equation}
  \begin{aligned}
    B^1_k & = \mathcal{T}^e_k \left( t_k^{n_k}, \Delta t_k \right)\\
    B^2_k & = \mathcal{T}^e_k \left( t_k^{n_k}, \frac{1}{2} \Delta t_k \right)\\
    B^3_k & = 
    \begin{cases}
        \mathcal{T}^e_k \left( t_k^{n_k}, \Delta t_k \right),& \text{if } n_k\;\text{even}\\
        \mathcal{T}^e_k \left( t_k^{n_k-1}, 2\Delta t_k \right),& \text{if } n_k\;\text{odd}.
    \end{cases}
  \end{aligned}
  \label{eq:lts_buffers}
\end{equation}
$B_k^1$ is used by $k$'s neighboring elements with equal time steps.
$B_k^2$ is only defined and used if $k$ has a neighboring element with a smaller time step.
$B_k^3$ is only defined and used if $k$ has a neighboring element with a larger time step.

\begin{figure}[htbp]
  \centerline{\includegraphics[width=\linewidth]{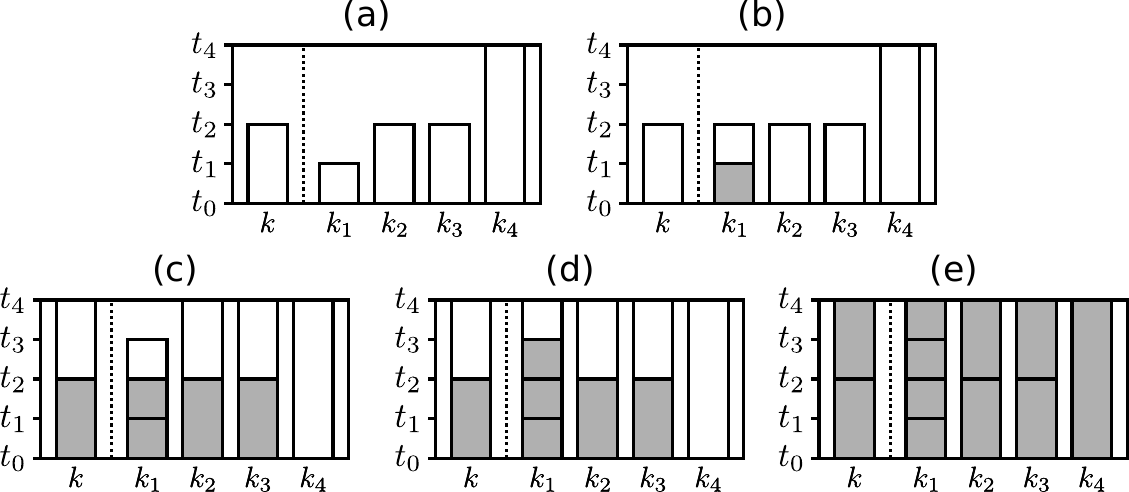}}
  \caption{Illustration of EDGE's next-generation clustered local time stepping scheme for an element $k$ and its four face-neighboring elements $k_1$, $k_2$, $k_3$ and $k_4$. White boxes indicate, that the element computed the time kernel and shared respective buffers (if any) with neighbors. Gray boxes indicate completed time steps.}
  \label{fig:lts_scheme}
\end{figure}
As an example, assume the situation illustrated in Fig.\,\ref{fig:lts_scheme}.
In the given case, $k$ has four neighboring elements $k_1$, $k_2$, $k_3$ and $k_4$.
The elements $k$, $k_2$ and $k_3$ advance with time step $2 \Delta t$.
$k_1$ advances with time step $\Delta t$, and $k_4$ advances with $4 \Delta t$.
In this example, we'll study how we store $k$'s data in $B^1_k$, $B^2_k$ and $B^3_k$, and how $k_1$, $k_2$, $k_3$ and $k_4$ use the buffers.
Note that elements $k_1$, $k_2$, $k_3$ and $k_4$ have to set their own buffers respectively, which is not part of the following considerations.
Initially, as shown in Fig.\,\ref{fig:lts_scheme}\,(a), all elements are at the same time level $t_0$ and compute time predictions.
Following Eq.\,\eqref{eq:lts_buffers}, element $k$ integrates its DOFs over the two time intervals $[t_0, t_0 + 2 \Delta t] = [t_0, t_2]$ and $[t_0, t_0 + \Delta t] = [t_0, t_1]$, and stores the result:
\begin{equation}
  \begin{aligned}
    B^1_k &= \mathcal{T}^e_k \left( t_0, 2 \Delta t \right)\\
    B^2_k &= \mathcal{T}^e_k \left( t_0,   \Delta t \right)\\
    B^3_k &= \mathcal{T}^e_k \left( t_0, 2 \Delta t \right).
  \end{aligned}
\end{equation}
Next, as shown in Fig.\,\ref{fig:lts_scheme}\,(b), only $k_1$ is allowed to complete its time step.
For this, it uses the time-integrated DOFs in $B^2_k$. 
Additionally, $k_1$ computes its next time prediction.
Now, as illustrated Fig.\,\ref{fig:lts_scheme}\,(c), elements $k$, $k_1$, $k_2$ and $k_3$ complete their respective time steps.
Here, $k_1$ uses the data in $B^1_k$ and $B^2_k$ to compute $k$'s time-integrated DOFs $\mathcal{T}^e_k(t_1, \Delta t) = B_k^1 - B_k^2$.
Elements $k_2$ and $k_3$ have the same time step as $k$ and use the data in $B_k^1$ directly.
Next, the elements $k$, $k_1$, $k_2$ and $k_3$ compute their new time predictions.
For $k$, following Eq.\,\eqref{eq:lts_buffers}, we update the buffers as follows:
\begin{equation}
  \begin{aligned}
    B^1_k &= \mathcal{T}^e_k \left( t_2, 2 \Delta t \right)\\
    B^2_k &= \mathcal{T}^e_k \left( t_2,   \Delta t \right)\\
    B^3_k &= \mathcal{T}^e_k \left( t_0, 2 \Delta t \right) + \mathcal{T}^e_k \left( t_2, 2 \Delta t \right) = \mathcal{T}^e_k \left( t_0, 4 \Delta t \right).
  \end{aligned}
\end{equation}
Buffer $B^2_k$ is then once again used for updating $k_1$, as shown in Fig.\,\ref{fig:lts_scheme}\,(d).
Further, in Fig.\,\ref{fig:lts_scheme}\,(e), $k_2$ and $k_3$ use $B^1_k$ for their update and $k_4$ uses $B^3_k$ for its update.

\subsection{Distributed Memory}
\label{ch:distributed_memory}
Our parallel implementation follows the ideas outlined in \cite{breuer2016petascale}.
A key difference is given in the communication scheme.
The work \cite{breuer2016petascale} sends time buffers or derivatives of elements which have face-neighbors in other partitions.
Our scheme relies solely on three time buffers, as outlined in Sec.\,\ref{ch:buffers}, which are used by face-neighboring elements in the shared memory domain.
For communication w.r.t. the distributed memory domain, we perform another compression step.
Instead of communicating the values of the entire buffers, we transform them to a face-local representation first.

Assume that an element $k$ face-neighbors another element $k_\text{neigh}$ in a different memory space.
In this case, $k$ has to share data, depending on the time stepping relation, from one or more of its buffers $B_k^1$, $B_k^2$, $B_k^3$ with $k_\text{neigh}$.
Following Eq.\,\ref{eq:neigh_elastic} and Eq.\ref{eq:neigh_anelastic}, $k_\text{neigh}$ requires this data for its surface kernel.
Element $k_\text{neigh}$ would now multiply the data with a flux matrix $\bar{F}_{j_k(i),h_k(i)} \in \mathbb{R}^{\mathcal{B} \times \mathcal{F}}$.
We harness this reduction from $9 \times \mathcal{B}$ to $9 \times \mathcal{F}$ values by conducting the flux-matrix multiplication as part of element $k$'s local update step and only sending the result of the matrix-product through the Message Passing Interface (MPI).
For a fifth order scheme, i.e., $\mathcal{B} = 35$ and $\mathcal{F}=15$, this procedure reduces the amount of communicated data if the element's buffers are only used by one or two face-neighboring elements which is a feasible assumption for a compact partitioning.
Note that this does not hold for the shared memory domain.
In this case, every tetrahedron has four face-neighbors if not at the boundary of the spatial domain.
Therefore, EDGE uses different approaches for communication in the shared and distributed memory domain.

As also done in \cite{breuer2016petascale}, we partition our meshes by assigning weights to the elements and faces.
Here, we assume that an element's computational effort solely depends on the time step of its respective cluster.
This means that elements of cluster $C_{1}$ are assigned the weight $2^{N_c-1}$, those of $C_2$ the weight $2^{N_c-2}$, ..., and those of cluster $C_{N_c}$ the weight $2^0 = 1$.
Additionally, we assign weights to faces based on the potential communication volume and frequency of the adjacent elements.
This information is then passed to a graph partitioner through the dual-graph of the mesh, where vertices in the dual-graph represent our mesh's elements, and vertices our mesh's faces. 
\begin{figure}[htbp]
  \centerline{\includegraphics[width=\linewidth]{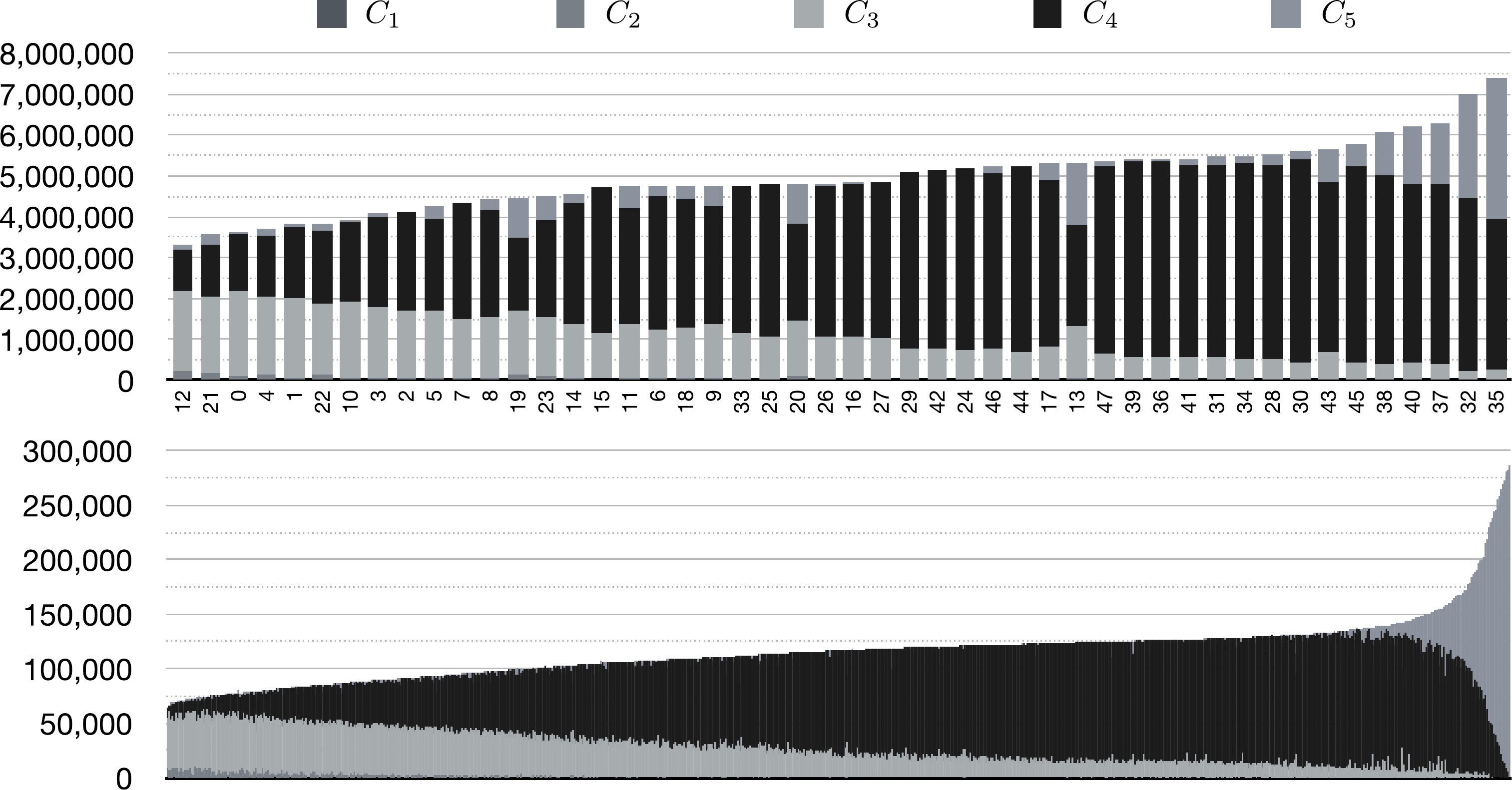}}
  \caption{Illustration of two partitionings for the studied 237,861,634-element 2014 $\text{M}_\text{w}$ 5.1 La Habra setting. (a) shows the used partitioning for 48 processes. (b) shows the used partitioning for 2048 processes. The partitions are ordered by their total number of elements. Colors in the stacked bar charts indicate the respective number of elements in the time clusters $C_1$, $C_2$, $C_3$, $C_4$ and $C_5$ (see Fig.\,\ref{fig:ts_density_la_habra}).}
  \label{fig:partitioning}
\end{figure}
As exemplary illustrated in Fig.\,\ref{fig:partitioning}, this procedure leads to a certain imbalance when considering the number of elements in a single partition.
This means that partitions with many elements belonging to large time step clusters have more elements in total.
Comparing the extremes of the 48-partition example, shown in Fig.\,\ref{fig:partitioning}\,(a), we obtain a $2.2\times$ difference, i.e., 3,311,441 tetrahedrons for the smallest partition and 7,378,861 elements for the largest one.
Doing the same for the 2048-partition example, shown in Fig.\,\ref{fig:partitioning}\,(b), we obtain only 64,569 elements for the smallest partition and 268,877 tetrahedrons for the largest one ($4.12\times$).
In practice, we always obtained usable partitionings when using $N_c \le 5$.
Future work may consider the implied memory footprint as an additional constraint in the partitioning.

\section{Production Pipeline}
\label{ch:pipeline}
EDGE's core solver offers powerful features through its support for viscoelastic attenuation, unstructured tetrahedral meshes, next-generation local time stepping, and fused simulations.
However, putting these features into production is a challenge from a modeling perspective.
This section introduces EDGE's new and rich preprocessing phase.

\begin{figure}[htbp]
  \centerline{\includegraphics[width=\linewidth]{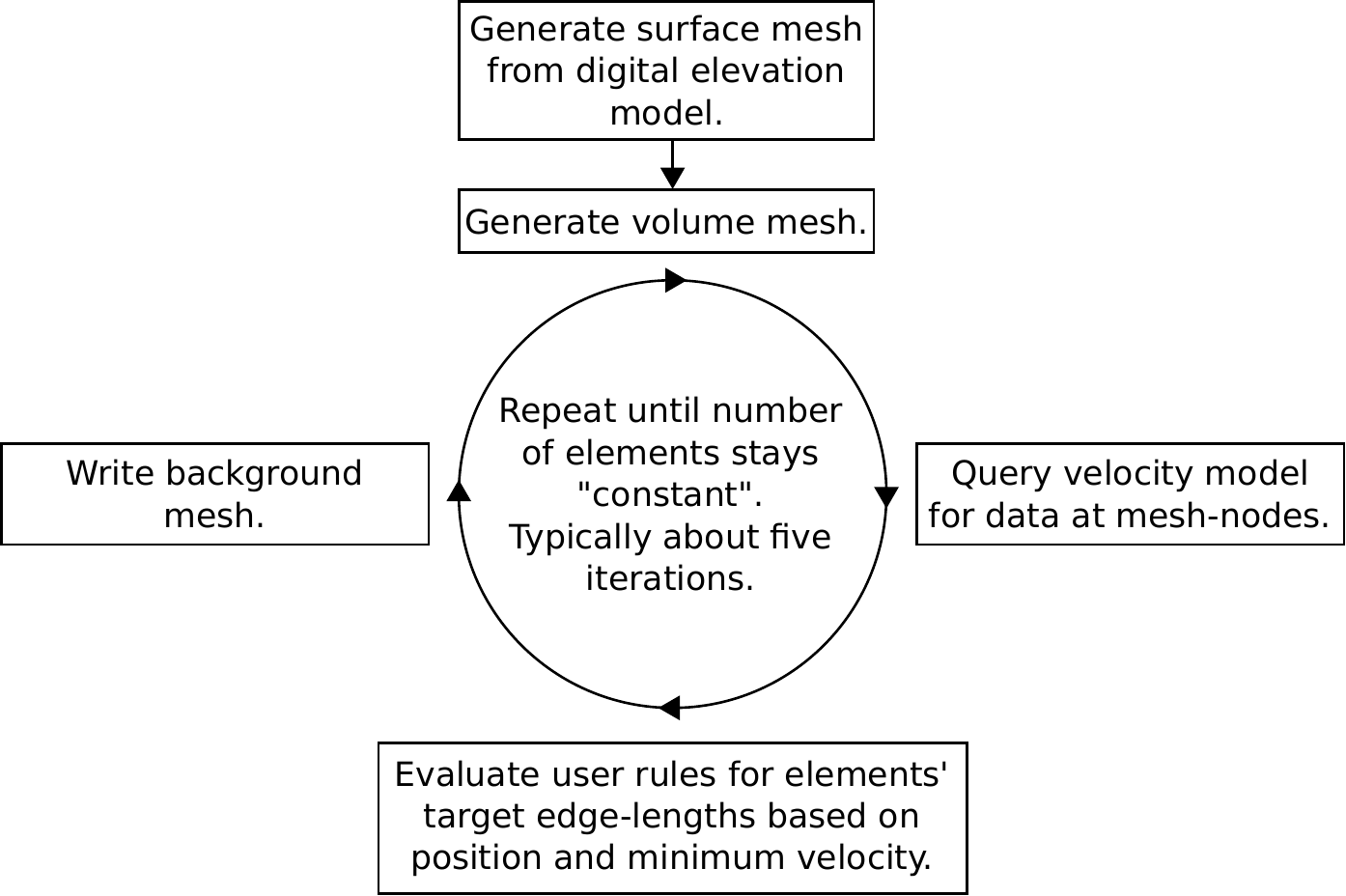}}
  \caption{Illustration of our velocity-aware meshing. We use EDGE's tool EDGEcut for the surface meshing, Gmsh \cite{geuzaine2009gmsh} as our volume mesher, and EDGE's tool EDGE-V to generate background meshes.}
  \label{fig:workflow_meshing}
\end{figure}
The preprocessing maximizes the algorithmic and computational performance of the core solver when running production workloads.
As illustrated in Fig.\,\ref{fig:workflow_meshing}, the first step of our preprocessing generates a problem-aware tetrahedral mesh.
Most importantly we provide target edge-lengths to the mesher based on the used seismic velocity model.
Here, we specify a certain number of elements per wave length through background meshes, resulting in highly refined low-velocity regions.

The next preprocessing step derives the local time stepping clusters and the optimal parameter $\lambda$, as discussed in Sec.\,\ref{ch:clustering}.
Next, as discussed in Sec.\,\ref{ch:distributed_memory}, we assemble weights for the elements, representing computation, and weights for the faces representing potential MPI-communication in the core solver.
This information is used to partition the mesh.
Then, we reorder the mesh based on the elements' partitions, time clusters, and finally by their role with respect to communication in the distributed memory parallelization.
The reordering greatly simplifies bookkeeping in terms of time clusters and elements involved in MPI-communication.
Further, it allows us to iterate linearly through blocks of our data in the time kernel, volume kernel, and local part of the surface kernel (see Eq.\ref{eq:update}).
The reordered mesh is written partition-wise to disk.
Additionally, we assemble a second file per partition which contains supporting data required by the core solver, e.g., the local MPI-communication structure or per-element seismic velocities.

At scale, each process reads its partition of the mesh and respective annotation file.
Both files contain all information required to initialize the core solver without additional MPI-communication.
In practice, our experiences with EDGE's new preprocessing phase have been far superior when compared to earlier approaches.
Our previous approaches had a now-obsolete online setup of the unstructured and partitioned mesh at scale which posed a major bottleneck in the past.
As part of High-F, we ran a simulation in the large domain (see Fig.\,\ref{fig:lahbra_map} and Fig.\,\ref{fig:verification}) using a velocity-adopted mesh with over 2.1 billion tetrahedrons.
This is a $4 \times$ increase in the number of elements, compared to the largest setting presented in \cite{id2703} (518 million elements).

\section{Simulations}
\label{ch:simulations}
\subsection{Frontera and General Setup}
The Frontera supercomputer is located at the Texas Advanced Computing Center and funded by the National Science Foundation.
The machine hosts a total of 8,368 Cascade Lake compute nodes.
Each of the nodes is equipped with two 28-core Intel Xeon Platinum 8280 processors and has 192\,GB of DDR4 main memory.
The processors have a base frequency of 2.7\,GHz.
Assuming a stable base frequency when running AVX512 instructions, we obtain a theoretical peak performance of 4.84\,FP32-TFLOPS.
The machine's interconnect is based on Mellanox HDR technology.
Full HDR (200\,Gb/s) connectivity is employed between switches and HDR100 (100\,Gb/s) between compute nodes.
Frontera ranks at position ten of the 06/21 TOP500-list with an achieved Linpack Performance of 23,516.4\,FP64-TFLOPS on 448,448 cores.

We conducted all simulations of this work on Frontera and in user operation.
All of our simulations used 32-bit floating point arithmetic and a fifth order instantiation of the studied solvers EDGE and SeisSol, i.e., $\mathcal{O}=5$.
Further, all of our simulations used three relaxation mechanisms for the anelastic part.
We studied the solver EDGE by running single and fused forward simulations.
In the latter case, we fused sixteen simulations, matching the processors' 16-value wide SIMD-instructions in single precision arithmetic (see Sec.\,\ref{ch:sparse_fused}).

We used EDGE's commit ae14203 and SeisSol's recent commit b76b440 for the reported performance results in this section.
In the considered version, SeisSol fails to compile when using its support for fused simulations.
Since this is a known issue \cite{pr2021fused} and since \cite{uphoff2020yet}'s results are limited to the elastic wave equations, we omit this part of the solver in this manuscript.

\subsection{LOH.3}
We use the Layer Over Halfspace benchmark 3 (LOH.3) \cite{day2003tests} to study EDGE's local time stepping accuracy and performance.
LOH.3 assumes a shear wave velocity of $v_s = 2000\,\text{m/s}$, a p-wave velocity of $v_p = 4000\,\text{m/s}$ and a density of $\rho = 2600\,\text{kg/m}^3$ in the upper $1000\,\text{m}$ of the model.
The model assumes $v_s=3464\,\text{m/s}$, $v_p=6000\,\text{m/s}$ and $\rho = 2700\,\text{kg/m}^3$ in the underlying halfspace.
The quality factors, assuming a constant Q-law, are set to $Q_s=40$ and $Q_p=120$ in the layer, and to $Q_s=69.3$ and $Q_p=155.9$ in the halfspace.
LOH.3 assumes a single point source at (0, 0, -2000\,m) to assess the accuracy of numerical solvers in a $\text{0-9}\,\text{s}$ time window.
Due to the simplicity of the setup, a quasi-analytical reference solution can be computed\footnote{In this work we use the LOH.3 reference solution from \url{http://sismowine.org/} which has been computed with the Axitra software.}.
Further, it allows us to compare EDGE's performance to the solver SeisSol which uses the LTS scheme described in \cite{breuer2016petascale}.

We used the software Gmsh \cite{geuzaine2009gmsh} to mesh the LOH.3 setup with an internal boundary in the Region Of Interest (ROI) at the material contrast at a depth of 1000m.
The shear waves are the slowest occurring waves in our model.
This means that the accuracy of the ADER-DG method is dictated by the spatial sampling of the shear waves if we assume a fixed configuration of the solvers, e.g., the order of convergence.
We obtained velocity-aware meshes by specifying a $1.732\,\times$ smaller characteristic edge length for the layer in the ROI which reflects the respective change in $v_s$.
Our first mesh has a total of 1,513,969 tetrahedrons and is the result of using a characteristic length of $200\,\text{m}$ for the layer and $346.6\,\text{m}$ for the halfspace in the ROI.
Additionally, we created a second, coarser mesh with a total of 743,066 tetrahedral elements for our performance comparisons, such that a single 28-core processor of Frontera is able to finish all studied configurations within a $48\,\text{h}$ runtime window.
We then used EDGE's preprocessing pipeline to prepare the input for EDGE's core solver.
Additionally, we converted the meshes to SeisSol's mesh format for our performance comparisons. 

\begin{figure}[t!]
  \centerline{\includegraphics[width=\linewidth]{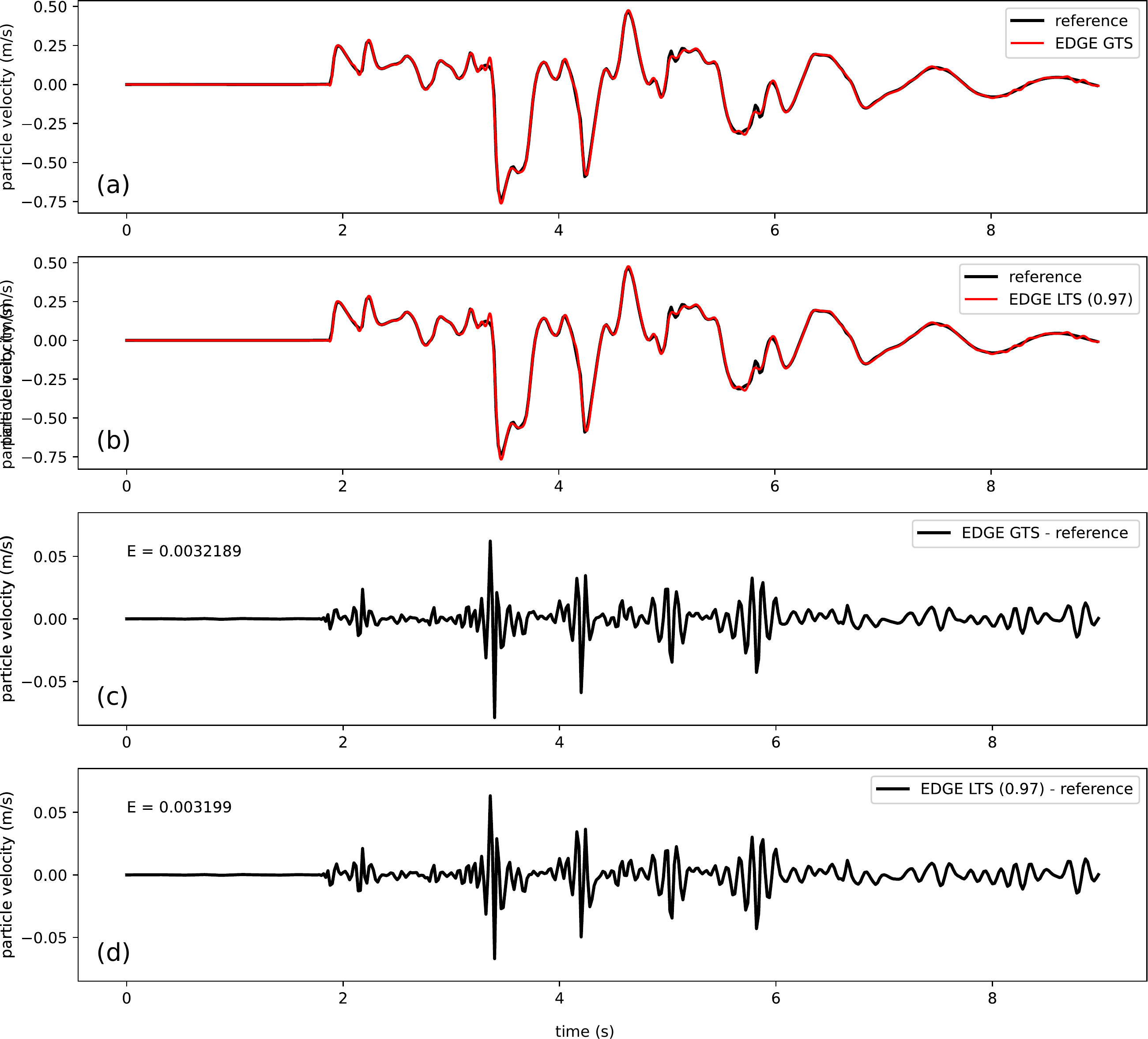}}
  \caption{Illustration of EDGE's Global Time Stepping (GTS) and Local Time Stepping (LTS) solutions for the ninth receiver of the Layer Over Halfspace 3 benchmark. (a) and (b) show the x-direction particle velocities (y-axis) over the nine simulated seconds (x-axis). Here, the reference solution is given through the black lines and EDGE's solutions through the slightly thinner red lines. Specifically, (a) shows EDGE's GTS-solution and (b) EDGE's LTS-solution for $\alpha = 0.97$. The remaining two plots, i.e., (c) and (d), show the respective differences w.r.t. the reference. For example, (d) shows the reference solution sample-wise subtracted from EDGE's LTS solution. }
  \label{fig:loh3_station}
\end{figure}

\paragraph{Accuracy}
Fig.\,\ref{fig:loh3_station} illustrates EDGE's accuracy when using different time stepping configurations for the solver.
Shown is the particle velocity in x-direction for the ninth receiver of the benchmark which is located at $(8647\,\text{m}, 5764\,\text{m}, 0)$.
We observe an excellent overall-fit when comparing EDGE's solutions (red) to the reference (black) in (a), (b).
Fig.\,\ref{fig:loh3_station} (c) and (d) show the respective differences w.r.t. the reference.
Once again, we observe almost identical results for the GTS solution in (c) and the LTS one in (d).
This is also confirmed quantitatively by the given seismogram misfits $E = \sum_{j=1}^{n_t} (s_j - s_j^\text{r})^2 / \sum_{j=1}^{n_t} (s_j^\text{r})^2$ \cite{kaser2007arbitrary}, where $n_t = 450$ is the number of time-samples of the seismogram, $s_j$ the respective particle velocity of EDGE's solution, and $s_j^\text{r}$ the velocity of the reference solution.

EDGE required 34,747 seconds to compute the GTS-solution on a single processor of Frontera.
In comparison, the LTS-run finished in only 5,788 seconds, reflecting a $6.0\,\times$ speedup.
This represents over $95\,\%$ of the theoretical $6.3\,\times$ speedup over GTS, offered by the increased algorithmic efficiency of the used grouped LTS-configuration.
Further, EDGE required 3,189 seconds to run the setup purely elastic, i.e., without viscoelastic attenuation, if using LTS.
Thus, EDGE's ``cost'' of anelasticity is about $1.8\times$ if using three relaxation mechanisms which is close to observations for SeisSol made in \cite{uphoff2016generating}.

\paragraph{Performance}
We use the smaller LOH.3 mesh with 743,066 tetrahedral elements to study and compare the performance of the two solvers EDGE and SeisSol on a single Frontera socket in detail.
\begin{table}[htbp]
  \caption{Performance of the two solvers EDGE and SeisSol on a single processor of Frontera when running the 743,066-element LOH.3 setting. The FP32-TFLOPS in hardware and the speedups over EDGE's single simulation GTS performance are given.}
  \begin{center}
    \begin{tabular}{|c|c|c|c|c|c|c|c|}
      \hline
      \textbf{Solver}
      &
      \textbf{Metric}
      &
      \multicolumn{2}{|c|}{\textbf{GTS}}
      &
      \multicolumn{2}{|c|}{\textbf{LTS (1.0)}}
      &
      \multicolumn{2}{|c|}{\textbf{LTS (0.8)}} \\
      && \textbf{\textit{1}} & \textbf{\textit{16}} & \textbf{\textit{1}} & \textbf{\textit{16}} & \textbf{\textit{1}} & \textbf{\textit{16}} \\
      \hline
      EDGE & TFLOPS   & 1.08 & 0.78$^\mathrm{a}$ & 1.01 & 0.74$^\mathrm{a}$ & 1.02 & 0.74$^\mathrm{a}$ \\
      \hline
      SeisSol & TFLOPS$^\mathrm{b}$ & 1.34&   -- & 1.09 &   -- &   -- & -- \\
      \hline
      EDGE & speedup   & 1.00 & 1.80 & 2.14 & 3.91 & 2.51 & 4.51 \\
      \hline
      SeisSol & speedup & 0.92 &   -- & 1.70 &   -- &   -- & -- \\
      \hline
      \multicolumn{8}{l}{$^{\mathrm{a}}$EDGE's fused simulations use sparse matrix kernels.}\\
      \multicolumn{8}{l}{$^{\mathrm{b}}$We report SeisSol's hardware performance as printed by the solver itself.}
    \end{tabular}
    \label{tab:speedups_flops_loh3}
  \end{center}
\end{table}
The first two rows of Tab.\,\ref{tab:speedups_flops_loh3} show the sustained single precision floating point performance of the two solvers in hardware.
We derived a total of 529,110 performed floating point operations per element update for EDGE's single forward simulations, only exploiting block-sparsity.
In contrast, the solver performed 212,688 floating point operations per simulation and element update when fusing sixteen simulations and exploiting all sparsity.
Thus, 59.8\% of the single simulation operations are zero-operations.
Additional analyses, studying respective ratios for the elastic wave equations, are given in \cite{heinecke2019tensor}.
For the solver SeisSol, we report the floating point performance as returned by the solver itself.
We observe that EDGE's floating point performance is between 20.9\% and 22.3\% of the processor's theoretical peak when running a single forward simulation.
The fused instantiation of the solver still reaches over 15\% of the theoretical peak performance in all settings.
The solver SeisSol reports a performance which is equivalent to a relative peak utilization of 27.7\% for GTS and 22.5\% for LTS.
Thus, it ``outperforms'' EDGE in this metric but is subject to a larger drop in performance when using LTS instead of GTS.
Note that raw floating point performance is not equivalent to time-to-solution. 
This becomes obvious when considering time-to-solution and accounting for increased algorithmic efficiencies in the next paragraph.
In fact, EDGE's lower floating point performance with a shorter time-to-solution indicates a superior implementation.

The last two rows of Tab.\,\ref{tab:speedups_flops_loh3} show the time-to-solution speedups of the two solvers relative to EDGE's GTS performance when running a single forward simulation.
First, by running sixteen fused simulations, we are able to increase the simulation throughput by $1.80\times$, even when using GTS.
Second, our next-generation LTS scheme is able to accelerate the solver by $2.14\times$ when using $N_c=3$ and $\lambda =1.0$ as discussed in Sec.\,\ref{ch:buffers} and shown in Fig.\,\ref{fig:ts_density_loh3}\,(a).
This reflects 94\,\% of the theoretical $2.28\times$-speedup and outperforms \cite{breuer2016petascale}'s scheme, implemented in SeisSol, by over $1.26\times$.
Combining both of EDGE's approaches, LTS with $\lambda=1.00$ and fused simulations, gives us a speedup of $3.91\times$.
Third, we can further enhance the LTS single simulation speedup to $2.51\times$ by using $\lambda=0.80$ as shown in Fig.\,\ref{fig:ts_density_loh3}\,(b).
Once again, also harnessing EDGE's fused simulations, we reach a final speedup of $4.51\,\times$, which outperforms the best performing single-simulation configuration of the solver SeisSol by $2.65\times$.

\subsection{2014 Mw 5.1 La Habra Earthquake}
Our setup for the 2014 $\text{M}_\text{w}$ 5.1 La Habra earthquake is an enhanced version of the High-F configuration (see Sec.\,\ref{ch:la_habra}).
We increased the complexity of the High-F setting in two crucial ways.
First, we incorporated topography information obtained from \cite{usgs3dep}.
Second, we reduced High-F's cut-off for the minimum shear wave velocity in the velocity model from 500\,m/s to 250\,m/s.
Both changes are challenging to realize in commonly used finite difference solvers.
Here, topography typically requires curved topography-aligned meshes while EDGE's preprocessing pipeline meshes topography explicitly.
The reduction of the minimum shear wave velocity requires an increased mesh resolution in the respective parts of the computational domain.
This is especially challenging if uniform grids with GTS are used since it leads to a costly oversampling elsewhere.
Specifically for the 2$\times$ reduction of the minimum shear wave velocity, one would have to invest $2^4 = 16 \times$ more compute resources for a regular mesh due to the three dimensions in space, and the time dimension.

We used EDGE's new preprocessing pipeline, outlined in Sec.\,\ref{ch:pipeline}, to generate a velocity-adapted and thus problem-aware mesh.
In the final step of the procedure, shown in Fig.\,\ref{fig:workflow_meshing}, we additionally increased the mesh resolution in regions with high velocity gradients.
The obtained final mesh had a total of 237,861,634 tetrahedral elements.
As discussed in Sec.\,\ref{ch:buffers} and visualized in Fig.\,\ref{fig:ts_density_la_habra}, we used $N_c=5$ time clusters and derived the parameter $\lambda = 0.81$.
This results in a theoretical speedup of $5.38\times$ for our clustered LTS scheme over GTS.

We strong-scaled the setup from 24 to 1,536 nodes of the Frontera machine when running a single simulation.
Further, we scaled sixteen fused simulations from 256 to 1,536 nodes.
\begin{figure}[htbp]
  \centerline{\includegraphics[width=\linewidth]{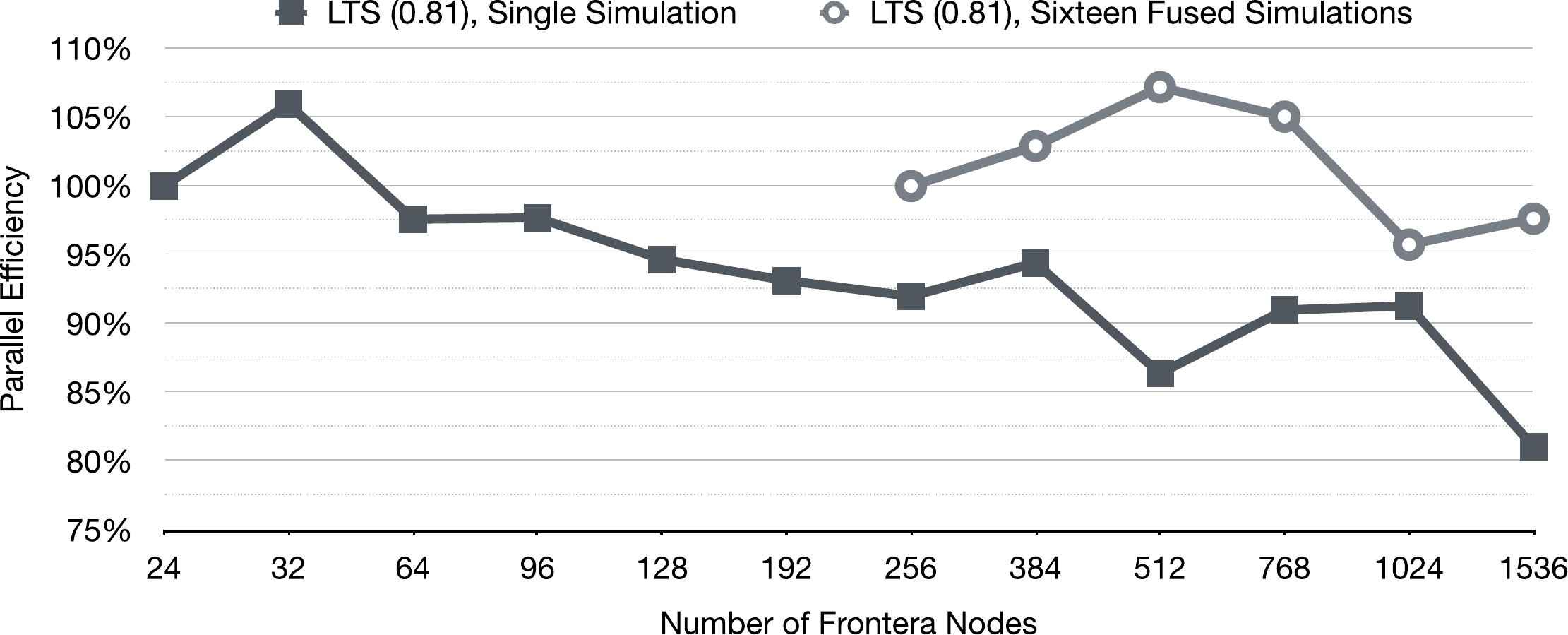}}
  \caption{Strong scaling studies of the solver EDGE for a ground motion simulation setup for the 2014 $\text{M}_\text{w}$ 5.1 La Habra earthquake. The used problem-aware tetrahedral mesh has 237,861,634 elements and includes mountain topography. Shown is the parallel efficiency of our next-generation local time stepping scheme on the Frontera supercomputer. We used the solver by running a single forward simulation and by fusing an ensemble of sixteen simulations.}
  \label{fig:scaling_la_habra}
\end{figure}
Fig.\,\ref{fig:scaling_la_habra} shows the sustained parallel efficiencies of the two scaling studies.
We observe that EDGE maintains a parallel efficiency of over 80\% in all configurations.
An explanation of the initially superlinear scaling might be given through NUMA-effects.
The single forward simulation on 24 nodes and the fused simulation on 256 nodes operated very close to the memory limit.
Here, data of a process might spill over to a different NUMA-domain despite our per-socket partitioning and pinning.
On 1,536 nodes we sustained a hardware performance of 2.25\,FP32-PFLOPS for a single forward simulation and 1.91\,FP32-PFLOPS when fusing sixteen simulations.
Respectively, we observe an 2.11$\times$ per-simulation speedup of the fused configuration.
Additionally, when harnessing all of EDGE's features on 1024 nodes, i.e., the presented next-generation local time stepping scheme and the solver's support for fused simulations, we obtain a 10.37$\times$ per-simulation speedup over a single forward simulation on 1,024 nodes using GTS.

\section{Conclusions and Availability}
This work presents a set of crucial extensions to the Arbitrary high-order DERivatives (ADER) Discontinuous Galerkin (DG) finite element software EDGE.
Our enhancements and developments cover the entire modeling and simulation spectrum.
First, we illustrated the solver's new support for the anelastic wave equations.
Second, we presented our next-generation local time stepping method and communication scheme for the ADER-DG method.
Third, we presented the software's new and rich preprocessing pipeline which allowed us to put the solver's advanced features into production.
The resulting software package outperforms the previous state-of-the-art LTS scheme, implemented in the SeisSol package, by 1.48$\times$.
When also harnessing EDGE's support for fused ensemble simulations, we obtained a combined speedup of 4.51$\times$ over using a single global time stepping simulation in EDGE.

We concluded our presentation by using EDGE for a demanding setup of the 2014 $\text{M}_\text{w}$ 5.1 La Habra earthquake.
In detail, we introduced mountain topography and derived a velocity-adapted and thus problem-aware mesh of the simulation region.
Strong scaling this setup to 1,536 nodes of the Frontera supercomputer, we reached a hardware-performance of 2.25\,FP32-PFLOPS for EDGE's new local time stepping scheme and when running a single forward simulation.
EDGE's fused simulation technology still reached a performance of 1.91\,FP32-PFLOPS, despite its reliance on sparse-matrix operations, leading to a higher algorithmic efficiency.
The solver's overall efficiency manifests when deploying all enhancements, presented in this manuscript, together with EDGE's support for fused simulations at scale.
Here, we obtained a 10.37$\times$ per-simulation speedup on 1,024 nodes of Frontera over the global time stepping performance of a single forward simulation.

EDGE is open source software and available from \url{https://dial3343.org} under the permissive 3-clause BSD license.
This includes all of EDGE's pre- and postprocessing tools used for this manuscript.
Further, we share a large amount of setups, scripts and data for EDGE at \url{https://opt.dial3343.org}.
EDGE exclusively relies on open source software in all of its components.
Intel MPI, the default and recommended MPI-implementation on the Frontera machine, is the only proprietary software which was used for running EDGE.

\bibliographystyle{IEEEtran}
\bibliography{IEEEabrv,references}

\end{document}